\documentstyle[twocolumn,epsfig]{mn}
\begin{document}

\title{Gravitational instability of discs with dissipative corona around supermassive black holes }

\author[F. Khajenabi and M. Shadmehri]{Fazeleh Khajenabi$^{1}$\thanks{E-mail:
fazeleh.khajenabi@ucd.ie (FK); } and Mohsen Shadmehri$^{2,3}$\thanks{E-mail:
mohsen.shadmehri@dcu.ie (MS)}\\
$^{1}$School of Mathematical Sciences, University College Dublin, Belfield, Dublin 4, Ireland\\
$^{2}$School of Mathematical Sciences, Dublin City University, Glasnevin, Dublin 9, Ireland\\
$^{3}$ Department of Physics, School of Science, Ferdowsi University, Mashhad, Iran}

\maketitle

\date{Received ______________ / Accepted _________________ }

\begin{abstract}
We study the dynamical structure of a self-gravitating disc with corona around a super-massive black hole. Assuming that the magneto-rotational-instability (MRI) responsible for generating the turbulent stresses inside the disc is also the source for a magnetically dominated corona, a  fraction of the power released when the disc matter accretes is transported to and dissipated in the corona. This has major effect on the structure of the disc and its gravitational (in)stability according to our analytical and self-consistent solutions. We determine the radius where the disc crosses the inner radius of gravitational instability and forms the first stars. Not only the location of this radius which may extend to very large distances from the central black hole, but also the mass of the first stars highly depends on the input parameters, notably the viscous coefficient, mass of the central object and the accretion rate. For  accretion discs around quasi-stellar objects (QSOs) and the Galactic center, we determine the self-gravitating radius and the mass of the first clumps. Comparing the cases with a corona and without a corona for  typical discs around QSOs or the Galactic center, when the viscosity coefficient is around 0.3, we show that the self-gravitating radius decreases by a factor of approximately 2, but the mass of the fragments increases with more or less the same factor. Existence of a corona implies a more gravitationally unstable disc according to our results. The effect of a corona on the instability of the disc is more effective when the viscosity coefficient increases.

\end{abstract}

\begin{keywords}
galaxies: active - black hole: physics - accretion discs
\end{keywords}
\section{Introduction}
Among various physical agents which play significant roles in accreting systems, self-gravity has a vital role in
accretion discs around supermassive black holes. According to the standard theory of accretion discs
(Shakura \& Sunyaev 1973), the outer parts of steady, geometrically thin, optically thick discs are prone to
self-gravity and they might be expected to fragment into stars (e.g., Shlosman \& Begelman 1987; Goodman 2003). While many authors proposed possible solutions to this problem (e.g., Goodman 2003), it seems such a gravitationally  unstable disc is a good explanation for the existence of the first starts in galactic centers or the formation of supermassive stars in quasar accretion discs (Goodman \& Tan 2004; Tan \& Blackman 2005; Levin  2007). Quasar spectra show that near the black hole there is high metallicity (Hamann et al. 2002), presumably because of massive star formation. Interestingly, young and massive stars are observed within $0.5$ pc of Sgr A$^\ast$ in our Galactic center (e.g., Lu et al. 2005; Paumard et al. 2006; Bender et al 2005). Based on the observation of the Galactic center, Levin \& Beloborodov (2003) proposed that a recent burst of star formation has occurred in a dense gaseous disc around Sgr A$^\ast$.  Also, the data of Paumard et al. (2006) strongly favour in situ star formation from dense disc in the central regions of the Galactic center. Star formation activities are seen in the nuclei of Seyfert galaxies (Terlevich 1996).

These observations have motivated many  authors to study star formation under extreme conditions in discs
around massive black holes at the center of  galaxies. The suggestion of star formation in the self-gravitating
parts of an accretion disc was first made by Kolykhalov \& Sunayev (1980). Shlosman \& Begelman (1989) discussed
in detail the conditions for star formation in the discs around black holes. Recently, star formation near to the Galactic Center has also been studied by Levin (2006) and Nayakshin (2006). Levin (2006) considers a disc that fragments into many stars of up to a few hundred solar masses and Nayakshin (2006) shows that the initial mass function for disc-born stars at the distances between $0.03$ and $0.3$ parsec from the supermassive black hole should be top-heavy. Collin \& Zahn (1999a, 1999b) have also studied formation of stars in AGN discs. Goodman \& Tan (2004) proposed that supermassive stars may form in quasar discs. According to their model, while the disc may fragment and each fragment may start with only a hundred solar mass, it seems likely they grow to a significant fraction of the disc mass, perhaps around $10^{5}$ solar mass. Just recently Nayakshin et al. (2007) studied star formation in the central parsec of our Galaxy using numerical simulations. However, they found that the masses of the first clumps are much smaller than $10^{5}$ solar mass which has been predicted by simple theoretical analysis (e.g., Goodman \& Tan 2004).

The observation of the UV bump of  AGNs (e.g., Malkan \& Sargent
1982) argues in favour of geometrically thin and optically thick
discs, possibly embedded in a hot X-ray emitting corona. In fact,
coexisting soft and hard X-ray components observed in ordinary AGNs
can not be well described by simple one-zone hot-flow models and
require composite disc-corona structure. Considering some common
features of the corona of a disc with a solar corona, interesting
theoretical investigations have also been presented in the frame of
a magnetized disc corona (Haardt \& Maraschi 1991; Field \& Rogers
1993; Nakamura \& Osaki 1993; Merloni \& Fabian 2002; Misra \& Taam
2002; Liu, Mineshige \& Shibata 2002; Merloni 2003). It has been
suggested that the corona consists of localized active regions. It
is likely that these are produced by magnetic fields in the disc
amplified through differential rotation. When the disc's magnetic
field builds up significantly, buoyancy forces the field out and
above the disc, giving rise to active regions of high magnetic
field. Misra \& Taam (2002) found that a significant part of the
accretion flow (or dissipation rate) can take place in the corona if
the scale height of the magnetic field is larger than that of the
disc. According to their study, optically thick discs with
dissipative coronas can provide an attractive explanations for the
origin of the soft spectral component observed in black hole X-ray
binary systems. R$\rm\acute{o}$$\rm\dot{z}$a$\rm\acute{n}$ska et al.
(1999) studied the vertical structure of a
radiation-pressure-dominated disc with a hot corona. They showed
that the presence of the corona modifies considerably the density
and the opacity of the disc surface layers and concluded that the
corona is not only essential from the point of view of the X-ray
spectra formation but helps to remove the problems with the disc
models for AGN. The disc-coronal
model for AGN has also been discussed  by Field \& Rogers (1993).

Merloni \& Fabian (2002) showed that, if angular momentum transport in the disc is due to
magnetic turbulent stresses, the magnetic energy density and
effective viscous stresses inside the disc are proportional to the
geometric mean of the gas pressure and the total pressure (gas plus radiation).
They discussed why energetically dominant coronae are ideal sites
for launching powerful jets/outflows, both MHD- and thermally
driven. Merloni (2003) presented   thin disc solutions accompanyied  by powerful,
magnetically dominated coronae and outflows, as models for black
holes accreting at super-Eddington rates.  Based on the assumption that the magnetorotational instability (MRI)
responsible for generating the turbulent stresses inside the discs
(Balbus \& Hawley 1991) is also the source for a magnetically
dominated corona, Merloni \&  Nayakshin (2006) studied the
limit-cycle instability in magnetized accretion discs. In these models, a fraction  of the gravitational energy
release is assumed to dissipate in a hot diffuse corona, above the
main body of the disc, away from the majority of the accreted mass.

Thus, one may conclude that corona is an important part of the accretion
processes around central black holes of AGNs. Possible
effects of a corona on its underlying disc have been studied using
different models. For example,  a dissipative corona may be able to thermally
stabilize the underlying disc (Misra \& Taam 2002). However, the possible effects of the corona on the gravitational
stability of the disc has not been studied to our knowledge. In this
study, we would like to address this question by a very simple
model: What is the effect of the corona on the gravitational stability of a
self-gravitating disc? We know that the corona may thermally
stabilises the disc, but can we expect gravitational stabilization
as a result of  the presence of the corona? Considering the growing interest in star
formation near our Galactic center and discs of AGNs, any possible
effects of the corona will have interesting implications in these
studies. In our approach, the disc structure equations are reduced to
simple algebraic relations in terms of average quantities of the
disc and corona.

\section{General formulation}
We assume that all angular momentum transport takes place in the disc and
the mass accretion rate $\dot{M}$ is constant with radius and time. However, the microphysics mechanism of the angular
 momentum transfer remains unknown.  Shakura \& Sunyaev (1973) replaced all the missing physics by a parameter $\alpha$.
 This approach has been widely used for studying the dynamics and structure of the accretion flows. A promising mechanism for
 driving the turbulence responsible for angular momentum and energy transport is the action of the magneto-rotational
 instability (MRI) that is expected to take place in such discs (Balbus \& Hawley 1991). But for a disc-corona system
 it is typically thought that the viscous stress, assumed to be magnetic in nature, transports angular momentum and
 initially randomise the gravitational binding energy near the midplane. The magnetized fluid elements, which are
 buoyant with respect to their surroundings, dissipate above the disc. Merloni \& Fabian (2002) and Merloni (2003)
 presented  very detailed discussions about dissipation in the corona and its relation with angular momentum transport
 in the disc itself. Our disc-corona model is developed along the
 line proposed by Merloni \& Fabian (2002) and Merloni (2003).

We  consider a more general prescription for the viscous stresses $\tau_{\rm r\phi}$ (Taam \& Lin 1984; Watarai \& Mineshige 2003; Merloni \&  Nayakshin 2006):
\begin{equation}
\tau_{\rm r\phi}=\alpha_{0} p^{1-\mu/2} p_{\rm gas}^{\mu/2},\label{eq:visg}
\end{equation}
where $\alpha_{0}$ and $0\leq\mu\leq2$ are constants. Also, $p$ is
the sum of the gas and radiation pressures. Phenomenological  models
generally assume that at each radius, a fraction $f$ of accretion
energy is released in the reconnected magnetic corona. Assuming that in
MRI-turbulence discs such a fraction $f$ of the binding energy is
transported from large to small depths by Poynting flux, Merloni \& Nayakshin
(2006) estimated the fraction $f$ as
\begin{equation}
f=\sqrt{2\alpha_{0} \beta^{\mu/2}},\label{eq:f}
\end{equation}
where $\beta$ is the ratio of gas pressure to the total pressure. So, in this model, this fraction $f$ is not a free parameter.

The rotation curve is dominated by a Newtonian point mass $M$, as
relativistic effects are only important at small radii. Although self-gravity may be important
for local stability at the outer radii of the disc, the rotation
angular velocity is not much affected because the disc is thin and
its mass is smaller than the central point mass (e.g., Goodman
2003). Thus, the rotational angular velocity of the disc is
Keplerian, i.e. $\Omega_{\rm K}=\sqrt{GM/R^{3}}$. Having equation
(\ref{eq:visg}) as a prescription for the viscous stresses and
equation (\ref{eq:f}) as an expression for the fraction of power
dissipated in the corona, we can write the basic equations of the
disc. Vertical hydrostatic equilibrium of the disc implies
\begin{equation}\label{eq: Zdirec}
\frac{p}{\Sigma}=\frac{\Omega_{\rm K}^{2}H}{2},
\end{equation}
and the azimuthal component of the equation of motion gives
\begin{equation}\label{eq:Phidirec}
8\pi\alpha_{0} H (p_{\rm gas})^{\mu/2} p^{(2-\mu)/2}= 3\Omega_{\rm K} \dot{M} J(R),
\end{equation}
where $J(R)=1-\sqrt{R_{\rm in}/R}$. Here, $R_{\rm in}$ denotes the
inner boundary of disc. Since we are interested in the regions of
the disc with radii much larger than $R_{\rm in}$, we can assume
$J\approx 1$. However, we will keep this factor in order to present
solutions of the disc structure as general as possible.

The energy equations is given by
\begin{equation}\label{eq:energy}
\sigma T_{\rm eff}^{4}=\frac{3}{8\pi}\Omega_{\rm K}^{2}\dot{M} J(R) (1-f),
\end{equation}
where as we discussed $f=\sqrt{2\alpha_{0}\beta^{\mu/2}}$. Also, we
can assume that the vertical transport of heat is by radiative
diffusion which implies the midplane and the surface temperatures
are related by
\begin{equation}
T = (\frac{3}{8}\kappa\Sigma)^{1/4} T_{\rm eff},
\end{equation}
where $\kappa$ is the opacity coefficient.

Now, equations (\ref{eq: Zdirec}), (\ref{eq:Phidirec}) and
(\ref{eq:energy}) are the main equations which enable us to find $p$
and $T$ as functions of $R$ and $\beta$ and the other input
parameters. Thus,
\begin{displaymath}
T=(\frac{4\sigma \Omega_{\rm K}}{3\kappa \alpha_{0}})^{-1/2}(\frac{16\pi^{2}\alpha_{0}^{2}ck_{\rm B}}{3\sigma \mu_{\rm m} m_{\rm H} \dot{M}^{2}\Omega_{\rm K}^{4}J^{2}})^{-1/3}
\end{displaymath}
\begin{equation}\label{eq:main1}
\times \frac{(1-\sqrt{2\alpha_{0}\beta^{\mu/2}})^{1/2}}{
(1-\beta)^{1/3}}\beta^{(4-\mu)/12},
\end{equation}
\begin{displaymath}
p=(\frac{4\sigma \Omega_{\rm K}}{3\kappa \alpha_{0}})^{-1/2}(\frac{16\pi^{2}\alpha_{0}^{2}ck_{\rm B}}{3\sigma \mu_{\rm m} m_{\rm H} \dot{M}^{2}\Omega_{\rm K}^{4}J^{2}})^{-2/3}
\end{displaymath}
\begin{equation}\label{eq:main2}
\times\frac{(1-\sqrt{2\alpha_{0}\beta^{\mu/2}})^{1/2}}{
(1-\beta)^{2/3}}\beta^{(8-5\mu)/12}.
\end{equation}
There is an algebraic equation for $\beta$ as follows
\begin{displaymath}
\frac{k_{\rm B}}{\mu_{\rm m} m_{\rm H}} (\frac{4\sigma\Omega_{\rm K}}{3\kappa \alpha_{0}})^{-3/2}
(\frac{8\pi \alpha_{0}}{3\Omega_{\rm K}^{2}\dot{M} J})^{2} (\frac{16\pi^{2}\alpha_{0}^{2}c k_{\rm B}}{3\sigma\mu_{\rm m} m_{\rm H} \dot{M}^{2}\Omega_{\rm K}^{4} J^{2}})^{-5/3}
\end{displaymath}
\begin{equation}\label{eq:beta}
 \times \frac{
(1-\sqrt{2\alpha_{0}\beta^{\mu/2}})^{3/2}}{(1-\beta)^{5/3}}\beta^{(8+\mu)/12}=1,
\end{equation}
where $\mu_{\rm m}$ is the mean particle mass in units of the hydrogen atom mass, $m_{\rm H}$. The other constants have their usual meanings.

In order to study behavior of our solutions, it is more convenient
to introduce dimensionless variables. For the central mass $M$,  we
introduce $M_8=M/(10^{8}M_{\odot})$ and for the radial distance $R$,
we have $r_3=R/(10^{3}R_{\rm S})$, where $R_{\rm S}=2GM/c^{2}$ is
the Schwarzschild radius. The mass accretion rate can be written as
\begin{equation}
\dot{M}=\frac{l_{\rm E}}{\epsilon}\frac{4\pi GM}{\kappa_{\rm e.s.}c}=\frac{l_{\rm E}}{\epsilon}\frac{L_{\rm E}}{c^{2}},
\end{equation}
where $l_{\rm E}=L/L_{\rm E}$ is the dimensionless disc luminosity relative to the Eddington limit and $\epsilon=L/(\dot{M}c^{2})$ is the radiative efficiency. Also, $\kappa_{\rm e.s.}\approx 0.04$ $\rm m^{2} kg^{-1}$ is the electron opacity. In our analysis, we will use the nondimensional factor  $l_{\rm E}/\epsilon$ as a free parameter so that by changing this parameter we can consider  appropriate values of the accretion rate. However, some authors introduce different forms for the accretion rate. For example, Nayakshin \& Cuadra (2005) who studied gravitational stability of the Galactic center, introduced $\dot{M}= (\dot{m}/\epsilon)(L_{\rm E}/c^{2})$  with $\epsilon \approx 0.06$ and $\dot{m}=0.03$ to $1$. These values, which are appropriate for the Galactic center, correspond to $l_{\rm E}/\epsilon \approx 0.5$ to $16.6$ in our notation. Also, for a central mass with mass $M=10^{8} M_{\odot}$, Goodman \& Tan (2004) proposed $l_{\rm E}/\epsilon = 10$. Thus, in our analysis, the chosen values of $l_{\rm E}/\epsilon = 1$  and $10$ are acceptable.

Now, we can rewrite our solutions as (in SI)
\begin{displaymath}
 \rho=2.76\times10^{-6} \alpha_{0}^{-1/2}\hat{\kappa}^{3/2} M_{8}^{-1/2}(\frac{l_{\rm E}}{\epsilon})^{2} J^{2} r_{3}^{-15/4}
\end{displaymath}
\begin{equation}
\times \beta^{(8-\mu)/4}(1-\beta)^{-2}(1-\sqrt{2\alpha_{0}\beta^{\mu/2}})^{3/2},\label{eq:rho}
\end{equation}

\begin{displaymath}
p=27.15\alpha_{0}^{-5/6}\hat{\kappa}^{1/2} M_{8}^{-5/6}(\frac{l_{\rm E}}{\epsilon})^{4/3} J^{4/3} r_{3}^{-13/4}
\end{displaymath}
\begin{equation}
 \times \beta^{(8-5\mu)/12}(1-\beta)^{-2/3}(1-\sqrt{2\alpha_{0}\beta^{\mu/2}})^{1/2},\label{eq:p}
\end{equation}

\begin{displaymath}
 \frac{H}{R}=4.67\times10^{-3}\alpha_{0}^{-1/6}\hat{\kappa}^{-1/2} M_{8}^{-1/6}(\frac{l_{\rm E}}{\epsilon})^{-1/3} J^{-1/3}
\end{displaymath}
\begin{equation}
 \times \beta^{-(8+\mu)/12}(1-\beta)^{2/3}(1-\sqrt{2\alpha_{0}\beta^{\mu/2}})^{-1/2},\label{eq:HR}
\end{equation}
and the ratio $\beta$ is obtained from nondimensional form of equation (\ref{eq:beta}), i.e.
\begin{displaymath}
 0.16 \alpha_{0}^{1/6}\hat{\kappa}^{3/2} M_{8}^{1/6}(\frac{l_{\rm E}}{\epsilon})^{4/3} J^{4/3} r_{3}^{-7/4}
\end{displaymath}
\begin{equation}
\times \beta^{(8+\mu)/12}(1-\beta)^{-5/3}(1-\sqrt{2\alpha_{0}\beta^{\mu/2}})^{3/2}=1,\label{eq:beta1}
\end{equation}
where $\hat{\kappa}=\kappa/\kappa_{\rm e.s.}$ and we assumed $\mu_{\rm m} \approx 0.6$. We can also calculate the surface density as
\begin{displaymath}
\Sigma = 1.27\times 10^{5} \alpha_{0}^{-2/3}\hat{\kappa} M_{8}^{1/3}(\frac{l_{\rm E}}{\epsilon})^{5/3} J^{5/3} r_{3}^{-11/4}
\end{displaymath}
\begin{equation}
\times \beta^{(4-\mu)/3}(1-\beta)^{-4/3}(1-\sqrt{2\alpha_{0}\beta^{\mu/2}}).
\end{equation}

Equations (\ref{eq:rho}), (\ref{eq:p}) and (\ref{eq:HR}) along with
equation (\ref{eq:beta1}) describe the structure of a disc with a
dissipative corona. However, the physical variables depend not only
on the radial distance but also on the ratio of the gas pressure to the
total pressure which can be calculated at each radius from algebraic
equation (\ref{eq:beta1}). For $\mu=1$, our solutions reduce to
what has been obtained by Merloni (2003). In the next section we
will analyze our solution, in particular the gravitational stability
of the disc.

\section{Analysis}
Since we are interested in gravitational stability of the disc, we
can approximate $J(R)\simeq 1$ for $R\gg R_{\rm in}$. Having the input parameters, we can solve equation (\ref{eq:beta1}) numerically at every  radius to obtain the ratio of the gas pressure to the total pressure. Then, equation (\ref{eq:f}) gives the fraction $f$ of gravitational power associated with the angular momentum transport that is transported vertically and dissipated in the corona. As the systems tends toward the gas pressure-dominated regime, the dissipated energy in the corona increases according to this equation. However, our
solutions  generally describe an inner radiation-dominated region
with an outer gas pressure-dominated region according to
equation (\ref{eq:beta1}). Moreover, the size of gas pressure-dominated regime increases, as the accretion rate decreases. These results are valid irrespective of the value of $\mu$ the exponent of the magnetic viscosity. For $\mu=1$, similar typical behaviors have already been obtained by Merloni \& Fabian (2002). We want to extend this analysis by studying the gravitational instability of the disc. In doing so, we should calculate the Toomre parameter of the disc.

Toomre (1964) showed that a rotating disc is subject to
gravitational instabilities when the $Q$-parameter becomes smaller
than a critical value, which is close to unity,
\begin{equation}
Q=\frac{c_{\rm s} \Omega}{\pi G \Sigma},
\end{equation}
where $c_{\rm s}$ is the sound speed inside the accretion disk and $\Omega = \Omega_{\rm K}$ is the angular velocity. So, the Toomre parameter
of our model becomes
\begin{displaymath}
Q = 44 \alpha_{0}^{1/2}\hat{\kappa}^{-3/2} M_{8}^{-3/2}(\frac{l_{\rm E}}{\epsilon})^{-2} J^{-2} r_{3}^{3/2}
\end{displaymath}
\begin{equation}
\times \beta^{-(8-\mu)/4}(1-\beta)^{2}(1-\sqrt{2\alpha_{0}\beta^{\mu/2}})^{-3/2}.\label{eq:Toomrem}
\end{equation}
This equation with algebraic equation (\ref{eq:beta1}) gives the Toomre parameter as a function of the radial distance. Generally, this parameter is much higher than unity in the inner parts of the disc which implies these regions are gravitationally stable and do not fragment. But the Toomre parameter decreases with increasing  radial distance so that $Q$ reaches the critical value of unity at a self-gravitating radius which we denote  by $R_{\rm  sg}$. Thus, all regions with $R>R_{\rm sg}$ are gravitational unstable and may fragment to clumps and cores.

\begin{table*}\label{t}
 \centering
 \begin{minipage}{140mm}
  \caption{The self-gravitating radius $R_{\rm sg}$ (in Schwarzschild radius $R_{\rm S}$) and the mass of the first clumps $M_{\rm frag}$ (in  solar mass) for $M=10^{8} M_{\odot}$, $\mu=1$ and $\hat{k}=1$.}
  \begin{tabular}{@{}lccccccccccc@{}}
  \hline
  without corona&&{$l_{\rm E}/\epsilon=1$}& &  & &$l_{\rm E}/\epsilon=10$&&&&\\
  \hline
            $\alpha_{0}$ &$0.03$ & & $0.3$ &
            &  $0.03$ &  & $0.3$ &
      \\

 \hline
 $R_{\rm sg}/(10^{3}R_{\rm S})$ & 0.5 &  & 3.8  &    & 0.75 &  & 2.3 & \\
  &  &  &   &    & &  & &   &    &  & \\
 $M_{\rm frag}/M_{\odot}$ & 993 &  & 17 &    & 14350 &  & 357 &  \\

\hline
with corona&&{$l_{\rm E}/\epsilon=1$}& &  & &$l_{\rm E}/\epsilon=10$&&&&\\
\hline
$\alpha_{0}$ &$0.03$ & & $0.3$ &
            &  $0.03$ &  & $0.3$ &
      \\

 \hline
 $R_{\rm sg}/(10^{3}R_{\rm S})$ & 0.4 &  & 1.7  &    & 0.66 &  & 1 & \\
  &  &  &   &    & &  & &   &    &  & \\
 $M_{\rm frag}/M_{\odot}$ & 1104 &  & 29.7 &    & 15530 &  & 577 &  \\

\hline

\hline

\end{tabular}
\end{minipage}
\end{table*}

\begin{table*}\label{tt}
 \centering
 \begin{minipage}{140mm}
  \caption{The same as Table 1, but  for $M=3\times 10^{6} M_{\odot}$.}
  \begin{tabular}{@{}lccccccccccc@{}}
  \hline
  without corona&&{$l_{\rm E}/\epsilon=1$}& &  & &$l_{\rm E}/\epsilon=10$&&&&\\
  \hline
            $\alpha_{0}$ &$0.03$ & & $0.3$ &
            &  $0.03$ &  & $0.3$ &
      \\

 \hline
 $R_{\rm sg}/(10^{3}R_{\rm S})$ & 470 &  & 6884  &    & 102&  & 1484 & \\
  &  &  &   &    & &  & &   &    &  & \\
 $M_{\rm frag}/M_{\odot}$ & 0.004 &  & $5\times 10^{-5}$ &    & 0.12 &  & 0.0016 &  \\

\hline
with corona&&{$l_{\rm E}/\epsilon=1$}& &  & &$l_{\rm E}/\epsilon=10$&&&&\\
\hline
$\alpha_{0}$ &$0.03$ & & $0.3$ &
            &  $0.03$ &  & $0.3$ &
      \\

 \hline
 $R_{\rm sg}/(10^{3}R_{\rm S})$ & 407 &  & 3275  &    & 88 &  & 704 & \\
  &  &  &   &    & &  & &   &    &  & \\
 $M_{\rm frag}/M_{\odot}$ & 0.0045 &  & $10^{-4}$ &    & 0.14 &  & 0.003 &  \\

\hline

\hline

\end{tabular}
\end{minipage}
\end{table*}

Different authors estimate the mass of fragments differently. Since
the disc is marginally unstable, the initial sizes and masses of
gravitationally bound fragments can be determined by Toomre's
dynamical instability (Toomre 1964). The most unstable wavelength for the $Q \sim
1$ disc is of order of the disc vertical scale height $H$ (Toomre
1964). Thus, the most unstable mode has radial wave number $k_{\rm
mu}=(QH)^{-1}$ and so the mass of a fragment at $R=R_{\rm sg}$
becomes
\begin{equation}
M_{\rm frag} \approx \Sigma (\frac{2\pi}{k_{\rm mu}})^{2}=4\pi^{2}\Sigma H^{2}.\label{eq:frag}
\end{equation}

\begin{figure}
\epsfig{figure=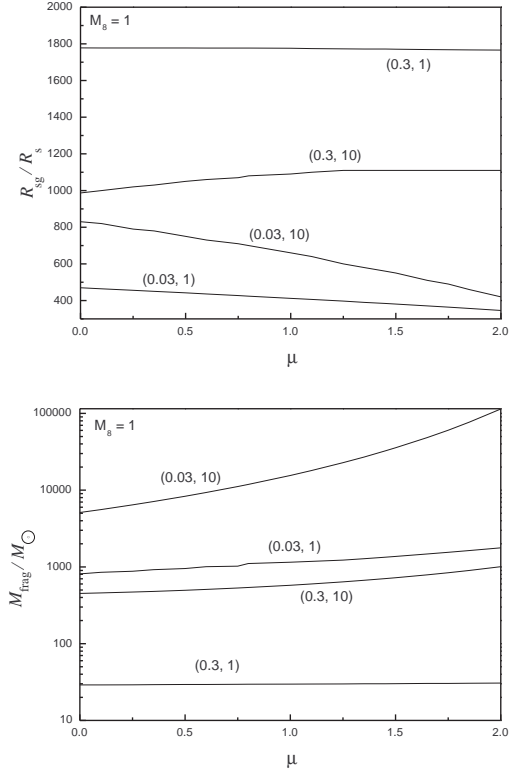,angle=0,width=8cm} \caption{The self-gravitating radius $R_{\rm sg}$ (in Schwarzschild radius $R_{\rm S}$) and the mass of the first clumps $M_{\rm frag}$ (in  solar mass) at this radius vs. the exponent $\mu$ of the viscosity prescription for a central black hole with mass $M=10^{8} M_{\odot}$ and $\hat{k}=1$. Each curve is marked by a pair ($\alpha_{0}$, $l_{\rm E}/\epsilon$) .}\label{fig:figure1}
\end{figure}

\begin{figure}
\epsfig{figure=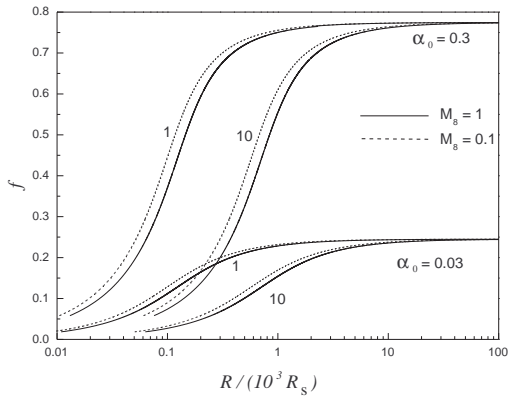,angle=0,width=8cm} \caption{The integral fraction $f$ of accretion power dissipated into the corona vs. radial location in the disc  with $\mu=1.5$ for $M=10^{8} M_{\odot}$ ({\it solid lines}) and $M=10^{7}M_{\odot}$ ({\it dashed lines}) central black hole with $\mu_{\rm m}=0.6$ and $\hat{k}=1$. The curves are labeled by the ratio $l_{\rm E}/\epsilon$. As this ratio increases, the fraction $f$ dissipated energy in the corona decreases, irrespective of the other input parameters. }\label{fig:figure2}
\end{figure}

\subsection{Parameter study}

Now, we can have a parameter study of the solutions. Figure \ref{fig:figure1} shows the self-gravitating radius $R_{\rm sg}$ (top) and the mass of the fragments (bottom) versus the exponent $\mu$ of the viscosity prescription for different input parameters. We fix the central mass ($M_{8}=1$) and the opacity ($\bar{\kappa}=1$), but vary the other input parameters. Each curve is labeled by the pair of the viscosity coefficient $\alpha_{0}$ and the accretion rate $l_{\rm E}/\epsilon$, as $(\alpha_{0}, l_{\rm E}/\epsilon)$. The top panel of Figure \ref{fig:figure1} shows that the disc of our system becomes self-gravitating at distances close to the central black hole. While for high accretion rate the self-gravitating radius $R_{\rm sg}$ depends on the exponent $\mu$, for lower accretion rate this radius is more or less independent of the exponent of the viscosity. For a fixed accretion rate, the self-gravitating radius $R_{\rm sg}$ increases with $\alpha_{0}$.

In fact, our viscosity prescription is different from the standard alpha model (Shakura \& Sunyaev 1973). In alpha model, Rice et al 2005 argue that the {\it maximum} viscosity in marginally stable self-gravitating discs is around 0.06. Unfortunately, there is not enough information about appropriate values of coefficient $\alpha_{0}$ for studying gravitational structure of discs. In analogy to previous studies of accretion discs based on the viscosity prescription of our model (e.g., Watarai \& Mineshige 2003), we are using a range of $0.03$ to $0.3$ for $\alpha_{0}$.

Using equation (\ref{eq:frag}), we can calculate the mass of the first fragments for the set of the above input parameters. For low accretion rate, the mass the fragments is also more or less independent of the exponent of $\mu$. But low values of $\alpha_{0}$ correspond to fragments with higher masses so that for pair of $(0.03, 10)$ the masses will be between $10^{4}$ and $10^{5}$ solar mass depending on the viscosity exponent $\mu$. However, as we mentioned, we think low values of $\alpha_{0}$ are not acceptable in self-gravitating discs. For a fixed accretion rate, the mass of the first clumps decreases with increasing $\alpha_{0}$. But as the accretion rate decreases, the mass of clumps decreases as well so that while for $(0.3, 10)$, we have $M_{\rm frag} \approx 600 M_{\odot}$ for $(0.3, 1)$ the mass of the fragments decreases approximately to  $M_{\rm frag} \approx 30 M_{\odot}$. It means for the above set of  input parameters, as the accretion rate decreases, fragments with lower masses are forming at larger distances from the central black hole. According to our results  not only the self-gravitating radius but also the mass of the first clumps show wide ranges of variations depending on the input parameters. In order to make easier comparison with other input parameters, Table 1 summarizes the self-gravitating radius and the mass of the first clumps for $M=10^{8} M_{\odot}$, $\mu=1$ and $\hat{k}=1$.  We will discuss the fraction of dissipated energy in the corona. But before that we can study the solutions for other set of the input parameters.

We found the self-gravitating radius and the mass of the fragments are not very sensitive to the exponent $\mu$ for lower mass of the central black hole. Thus, we summarize our results for  $M=3\times 10^{6} M_{\odot}$, $\mu=1$ and $\hat{k}=1$ in Table 2 which is appropriate for modeling the Galactic center. Clearly, when the mass of the central black hole decreases, the disc is more gravitationally stable as $R_{\rm sg}$ is much larger than in Figure \ref{fig:figure1} and the mass of the fragments are much smaller. In this case, also the typical behaviors of the solutions are similar to the case with $M=10^{8} M_{\odot}$. We see that as the viscosity coefficient $\alpha_{0}$ increases, not only does the self-gravitating radius increase, but the mass of the fragments decrease. Also, as the accretion rate increases, the self-gravitating radius decreases and the mass of the fragments increases as well.

Goodman \& Tan (2003) also estimated the mass of first clumps in quasar discs. Clearly, the disc of their model is without corona. Assuming that the viscosity is proportional to gas pressure, they found $R_{\rm sg}\approx 1700 R_{\rm s}$ to $R_{\rm sg}\approx 2700 R_{\rm s}$ with a few hundred solar mass for the fragments. Another point is that their model is radiation pressure dominated.  In another study, Nayakshin (2006) studied star formation near to our Galactic center. The mass of the first stars of this model is a few solar mass which may increase because of the subsequent accretion. However, our model shows  wider ranges for the self-gravitating radius and the mass of the fragments.  Moreover, when we reach the self-gravitating radius, the disc of our model is not necessarily radiation pressure dominated. In fact, we found the transition to the self-gravitating regime happens at radii where generally, the gas pressure, if it is not dominant, is at least comparable to the radiation pressure.

We can calculate the fraction $f$ of dissipated energy in the corona. Instead of integrating the fraction $f$ over all the disc, for our purposes it is sufficient  to plot this fraction as a function of the radial distance. Figure \ref{fig:figure2} shows the typical behavior of $f$ for our sets of the input parameters. Note that since we are not interested in the very inner regions of the disc, this plot is not accurate near these regions as we approximated $J\approx 1$. This figure shows interesting behavior. First of all, as the viscosity coefficient increases, the fraction $f$ increases as well. We also showed the self-gravitating radius increases and the mass of the fragments decreases with increasing $\alpha_{0}$. It means the disc is more  gravitationally stable as $\alpha_0$ increases or correspondingly, the amount of the dissipated energy in the corona increases. In order to confirm this result, we should check the behavior of $f$ and the disc for the other input parameters.

Consider Figure \ref{fig:figure2} for a fixed coefficient $\alpha_{0}$. We see that as the accretion rate decreases, the fraction $f$ increases which means more energy dissipates in the corona. Also, as the mass of the central black hole decreases, the fraction $f$ increases for a fixed accretion rate. For these cases, we showed that the disc becomes self-gravitating at larger radii with lower masses for the fragments, as the mass of the central black hole decreases (compare with Figure \ref{fig:figure1} and Table 1).

\subsection{The effect of corona}

How can we understand the above results based on the existence of the corona? For addressing this question, we can either compare our results directly with previous authors or with our calculations for the standard case (no corona). Although we briefly mentioned previous studies in this field, e.g.  Goodman \& Tan (2004) and Nayakshin (2006), we can not make direct comparisons with those studies because our basic assumptions for the corona and viscosity prescription are different. We think a straightforward way to understand the effect of the corona on the gravitational instability of the disc is direct comparison with our solutions, but without a corona.

Obviously, a certain amount of the generated energy in the disc transports in the corona in our model. This leads to a cooler disc in comparison  with a disc without corona. One may conclude that since the sound speed and consequently the Toomre parameter decrease, a disc with corona will be more unstable gravitationally. This result is correct as long as the surface density does not change or increase. In our model, the surface density is also lower than in the case of a disc without corona. So, this qualitative consideration does not help us to determine the typical behavior of the Toomre parameter as a result of existence of the corona.

\begin{figure}
\epsfig{figure=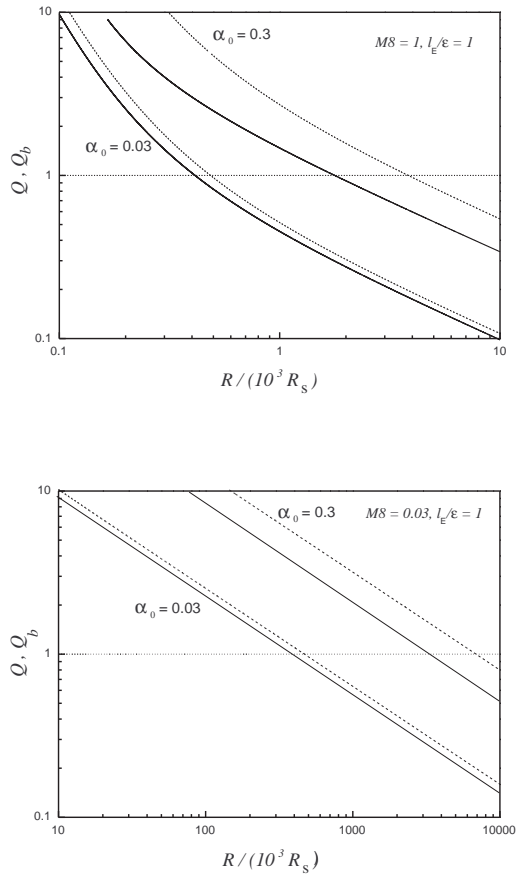,angle=0,width=8cm} \caption{The Toomre parameter for a disc with a corona ({\it solid lines}) and without a corona ({\it dashed lines})  vs. radial location in the disc  with $\mu=1.0$, $\mu_{\rm m}=0.6$ and $\hat{k}=1$ for $M=10^{8} M_{\odot}$ (top) and $M=3\times10^{6} M_{\odot}$ (bottom). }\label{fig:figure3}
\end{figure}

Now, we can simplify equation (\ref{eq:Toomrem}) by calculating $r_{3}$ from equation (\ref{eq:beta1}),
\begin{displaymath}
Q=9.14\alpha_{0}^{9/14}\hat{\kappa}^{-3/14}M_{8}^{-19/14}(\frac{l_{\rm E}}{\epsilon})^{-6/7}
\end{displaymath}
\begin{equation}
\times\beta^{(9\mu-40)/28}(1-\beta)^{4/7}(1-\sqrt{2\alpha_{0}\beta^{\mu/2}})^{-3/14}.
\end{equation}
Also, the Toomre parameter $Q_{\rm b}$ of the disc without a corona becomes
\begin{displaymath}
Q_{\rm b}=9.14\alpha_{0}^{9/14}\hat{\kappa}^{-3/14}M_{8}^{-19/14}(\frac{l_{\rm E}}{\epsilon})^{-6/7}
\end{displaymath}
\begin{equation}
\times\beta^{(9\mu-40)/28}(1-\beta)^{4/7}.
\end{equation}
Having all the input parameters fixed at each $\beta$ there is a relation between Toomre parameters of discs with (and without) corona:
\begin{equation}
\frac{Q}{Q_{\rm b}}=(1-\sqrt{2\alpha_{0}\beta^{\mu/2}})^{-3/14}.
\end{equation}
Since the expression inside the above parenthesis is always lower than unity, the above equation implies $Q>Q_{\rm b}$ for each $\beta$. On the other hand, numerical solutions of equation (\ref{eq:beta1}) for discs with (and without) corona show that $r_{\rm 3,b}>r_{3}$ for a fixed $\beta$. It means that the disc without corona reaches to a fixed $\beta$ at larger radius in comparison with a disc with corona. When we plot the Toomre parameter versus the distance from the central black hole, we have always $Q_{\rm b} > Q$ at a fixed radius. Thus, the Toomre parameter $Q$ of the disc with corona reaches  unity at smaller radius in comparison with a disc without corona. In other words, the self-gravitating radius of a disc with corona is smaller than disc without corona. We can confirm this general result by direct numerical solution of the equations. Figure \ref{fig:figure3} shows typical behavior of the Toomre parameter for the cases with and without a corona. Also,  Tables 1 and 2 show the self-gravitating radius and the mass of the fragments of the discs with and without corona. In the last subsection we discussed  general properties of  discs with corona. Now, we can compare the gravitational instability of the discs with and without a corona.

By comparing the Tables, we can simply conclude that as the mass of the central black hole increases, the discs become more self-gravitating irrespective of the existence of corona. In fact, while the self-gravitating radius $R_{\rm sg}$ decreases with the mass of the central object, the estimated mass of the clumps increases. These Tables confirm our above analytical expectation that in discs with a corona the Toomre parameter is lower than the same disc but without a corona. For example, Table 1 compares self-gravitating radius and the mass of the clumps in discs with and without corona, but both with $M=10^{8} M_{\odot}$. We see that not only  the self-gravitating disc decreases because of the corona, but also the mass of the clumps increases. But the effect of a corona on the instability of the disc is more effective for $\alpha_{0}=0.3$ comparing to $\alpha_{0}=0.03$. It means as the viscosity coefficient $\alpha_{0}$ increases, the corona makes the disc more gravitationally unstable. In fact, in our model, the fraction of the dissipated energy into the corona is directly proportional to the viscosity coefficient. So, as this parameter increases, since more energy dissipated in the corona, we see the effect of the corona more effectively. Comparing the cases with a corona and without a corona for $\alpha_{0}=0.3$, we see that $R_{\rm sg}$ decreases by  a factor of approximately 2, but $M_{\rm frag}$ increases with more or less the same factor because of the existence of the corona.

The above behaviors are also seen in Table 2, but the mass of the first clumps are much smaller in the case of $M=3\times 10^{6} M_{\odot}$ which corresponds to Sgr A$^\ast$, comparing to the other case. Also, in this case, the self-gravitating radius is much larger. However, evidently corona increases the mass of the first clumps and reduces the self-gravitating radius. Massive stars were found at a distance of order $0.1-0.3$ pc from Sgr A$^\ast$. For this central back hole, we have $0.1$ pc $\sim$ $3\times 10^{5} R_{\rm S}$. According to Table 2, the self-gravitating radius of our models is within the range $0.03$ to $2.3$ pc, depending on the existence of the corona and the input parameters. For example, consider a disc without corona with $l_{\rm E}/\epsilon = 10$ and $\alpha_{0}=0.3$, this table gives $R_{\rm sg} \sim 0.5$ pc. But when we consider the same disc but with corona, the self-gravitating radius decreases to $R_{\rm sg} \sim 0.25$ pc. This new location of the first clumps is consistent to typical distance of the observed massive stars in the Galactic center. Nayakshin \& Cuadra (2005) found the self-gravitating radius for their standard disc is around $0.03$ pc.  We note that their standard disc model is not similar to our standard model, but  since without a corona, we calculate a larger radial distance for the first clumps, if we modify the model Nayakshin \& Cuadra (2005) to include a corona, then the self-gravitating radius decreases.  If we consider higher values for the viscosity coefficient, the discrepancy between the cases with and without corona would be more significant.

\section{Discussion}
We have studied the effect of corona on the gravitational stability
of the disc by assuming that the hot corona and the disc coexist at
the same radius from the central black hole. In our disc-corona
model almost all the mass exists in the disc, but the dissipated
energy in the corona depends on the input parameters such as mass
accretion rate, mass of the central black hole and the exponent of
the viscosity prescription. Our study shows that corona has a destabilizing effect on the gravitational stability of the disc.  However, our disc and corona system is still able to explain the formation of rings of massive stars  observed in the Galactic center and in the nucleus of M31 (Davies et al. 2006). But star formation in self-gravitating disc with corona may have significant differences with discs without corona. Considering  results of our study, one can try to determine the initial mass function for disc-born first clumps at the inner region of a self-gravitating disc with corona around a super massive black hole, though Nayakshin (2006) has already predicted a top-heavy initial mass function for the first starts at the distances between $0.03$ and $0.3$ parsec from our Galactic center. We think that the estimated supermassive mass of the fragments according to Goodman \& Tan (2004) will change because of the destabilizing effect of corona which transports part of the generated energy inside the disc to the corona.

As we discussed, there are significant observational evidences in favor of star formation activities in accretion disc of AGNs. The mechanism of accretion is an important part of any theory for accretion discs. Most of the analytical and numerical simulations show that MRI is the most significant process in accretion discs. Almost all models of star formation in accretion discs use MRI as the main physical process of accretion. Does MRI have other consequences for the structure of the disc? As we discussed in the introduction section, the idea of a hot corona lying above the AGN accretion disc has been mainly developed in order to explain substantial emission in   the X-ray range. There are interesting analytical and numerical studies for understanding processes of corona formation (e.g., Heyvaerts \& Priest 1989; Miller \& Stone 2000). Interestingly, current models of disc and corona systems replace the fraction of accretion power transferred from the disc to the corona with a Poynting flux quantity estimated from a mean field buoyant velocity and an equipartition, mean field magnetic energy density. In other words, MRI may lead to corona formation as well, even though all the details of the process itself are not yet understood. Heyvaerts \& Priest (1989) have shown, loops and arcades can form also in AGN accretion discs; these structures, connecting remote points of the disc itself, can convert disc kinetic energy into magnetic energy and subsequently dissipate it by emitting the observed spectrum. Numerical simulations indicate that turbulent fluctuations in a vertically stratified disc are capable of driving the magnetogravitational modes of the Parker instability (e.g., Miller \& Stone 2000). It seems there is a {\it coexistence} between MRI and corona, at least in AGNs which its X-ray emission can be explained based on the existence of corona. Since our model is also based on this idea that MRI inside of the disc is also the source for a magnetically dominated corona, we think, this model is applicable to such AGNs, if the accretion disc itself is massive enough.

One should note that we have not followed the evolution of the first clumps after formation. One of the main important factors is accretion onto these new fragments. In fact, although the typical mass of the first clumps corresponding to the Galactic center is low comparing to the cases with higher central mass, we think, the accretion onto first clumps is able to increase the mass of first clumps significantly. This accretion process is believed to be similar to the growth of terrestrial planets in a planetesimal disc. In order to estimate, the amount of the accreted mass onto the first clump, we should calculate the Hill's radius which is directly proportional to the mass of the clump itself at the time of formation. Since corona increases the mass of the first clumps, the corresponding Hill's radius and subsequently the rate of the accretion onto it will increases. It means the final mass of the first clumps is even much higher because of the corona and subsequent enhanced accretion rate comparing to a case without corona.

Star formation feedback is indeed able to slow down disc
fragmentation as suggested by several authors, but since our goal was only to study the possible effects of the corona on the gravitational instability of the disc, we did not included star formation feedback in our model. Also, we have not studied the evolution of the clumps after formation. In fact, we showed that corona may have significant effects on the star formation in AGN disc or even near our Galactic center which should be considered in any successful theoretical model. Since our model for disc corona system is a phenomenological model, more careful description of the disc-corona transition is necessary. This transition zone may have an important effect on the emergent radiation flux. On the other hand, we did not investigate the structure of the corona itself, because in our simple model there is only energy exchange between corona and the underlying optically thick disc. In future works, it would be interesting to study gravitational stability of a system of disc and corona, in which not only the corona itself can transport the angular momentum, but also there are exchange of mass, energy and angular momentum between disc and corona.\\

\section*{Acknowledgments}

We gratefully acknowledge Peter Duffy for his support and encouragement, and the anonymous referee for remarks and suggestions leading to significant improvement of the paper and clarification of the presented results. We also thank B. Reville for reading the manuscript and making  useful comments. F. K. is grateful for Ad Astra PhD Scholarship of University College of Dublin. The research of M. S. was funded under the Programme for Research in
Third Level Institutions (PRTLI) administered by the Irish Higher
Education Authority under the National Development Plan and with partial
support from the European Regional Development Fund.

{}

\end{document}